\documentclass[pra,aps,showpacs,floatfix,preprint]{revtex4-1}
\usepackage{color}

\RequirePackage{amsfonts}

\usepackage{colordvi}

\usepackage{graphicx,amssymb,epsfig,amsmath}
\usepackage{epstopdf}

\newcommand{\be}{\begin{equation}}
\newcommand{\ee}{\end{equation}}
\newcommand{\bea}{\begin{eqnarray}}
\newcommand{\eea}{\end{eqnarray}}

\begin{document}
\title{Many-body entropies, correlations, and emergence of statistical relaxation in interaction quench dynamics of ultracold bosons}
\author{Axel U. J. Lode}
\email{axel.lode@unibas.ch} 
\affiliation{Department of Physics, University of Basel, Klingelbergstrasse 82,
CH-4056 Basel, Switzerland}
\author{Barnali Chakrabarti}
\affiliation{Department of Physics, Presidency University, 86/1 College Street,
Kolkata 700083, India}
\author{Venkata K. B. Kota}
\affiliation{Physical Research Laboratory, Ahmedabad, 380009, India }
\date{\today}

\begin{abstract}

We study the quantum many-body dynamics and the entropy production triggered by an interaction quench in a system of $N=10$ interacting identical bosons in an external one-dimensional harmonic trap. 
The multiconfigurational time-dependent Hartree method for bosons (MCTDHB) is used for solving
the time-dependent Schr\"odinger equation at a high level of accuracy. We consider many-body entropy measures such as the Shannon information entropy, number of principal components, and occupation entropy that are computed from the time-dependent many-body basis set used in MCTDHB. These measures quantify relevant physical features such as irregular or chaotic dynamics, statistical relaxation and thermalization. We monitor the entropy measures as a
function of time and assess how they depend on the interaction strength. For larger interaction strength, the many-body information entropy approaches the value predicted for the Gaussian orthogonal ensemble of
random matrices and implies statistical relaxation. The basis states of
MCTDHB are explicitly time-dependent and optimized by the variational principle
in a way that minimizes the number of significantly contributing ones. It is therefore a non-trivial fact that statistical relaxation prevails in MCTDHB computations. Moreover, we demonstrate a fundamental connection between the production of entropy, the build-up of correlations and loss of coherence in the
system. Since the coherence and correlations are experimentally accessible, their present connection to many-body entropies can be scrutinized to detect statistical relaxation. Our results are the first ones obtained for thermalization of finite quantum systems using an optimized time-dependent and genuinely many-body basis set.

\end{abstract}


\maketitle

\section{Introduction} 

The onset of thermalization in an isolated quantum system with a finite number of
interacting particles is an important issue in theoretical physics in
recent days \cite{deut, sred, rigo, polk, marc, zhua, rigol, beug, rigo1,
chav,kota,banu, rigo2, sant1, flam}. Experimental progress with various interacting quantum
systems has further corroborated the interest \cite{grei, trot}. These experiments reveal that the question of how entropy is
produced in a quantum system is one of the basic outstanding problems of
many-body physics \cite{rigo1, chav,kota,banu, rigo2, sant1, flam, grei, trot}.
A necessary condition for thermalization is statistical relaxation in 
various observables of a system to some kind of equilibrium
\cite{sant2,chir}. In isolated dynamical quantum systems of interacting
particles, statistical relaxation is related to chaos in
the energy spectra \cite{xxx, sant3} and here the inter-particle interaction
plays an important role. Chaos in quantum systems is in turn defined by the statistics of eigenstates. For time reversal and rotationally invariant
systems these follow the predictions of the Gaussian orthogonal ensemble of random
matrices (GOE), see Refs.~\cite{sred,rigo,sant2,sant3}.

The eigenstate thermalization hypothesis (ETH, put forward in Ref.~\cite{sred}) was so far only 
tested using models that employ a fixed basis set related to noninteracting particles. For example,
calculations have been done for one-dimensional spin-$\frac{1}{2}$ systems
with nearest-neighbor and next-nearest-neighbor coupling as well as for gapped
systems of hard-core bosons \cite{rigo1,sant2,sant3}. Statistical relaxation
was found, for large enough interaction strength, in the Shannon information entropy and the number of principal components.
Similarly, the manifestation of classical chaos in the statistics
of quantum energy levels and the confirmation of random matrix fluctuations in
molecular spectra has been established in Refs. \cite{thz1,thz2,thz3}.

In the present work, we consider the quantum many-body dynamics of ten bosons confined in a one-dimensional
harmonic trap that interact with a contact interaction potential. The one-dimensional regime is achieved experimentally in optical and magnetic traps with tight transverse confinement and frozen radial degrees of freedom. Quantum many-body effects are more important in reduced dimensional and interacting systems, because the
competition between statistical properties and quantum fluctuations is enhanced
in them. Experimentally, one-dimensional harmonically trapped quantum degenerate systems have
been realized; see, for instance, Ref.~\cite{gorl}. It is hence of fundamental both theoretical and experimental interest to understand the time-evolution of
entropy and onset of statistical relaxation in the quantum many-body dynamics of
the ultracold Bose gas in one dimension.  

We consider quench dynamics, where our $10$-boson system is perturbed by abruptly switching on the two-body part of the Hamiltonian. Increasing the positive prefactor, $\lambda_0$, of this two-body part of the Hamiltonian moves the system further and further away from integrability. We refer to $\lambda_0$ as strength of the inter-particle interactions. The process of statistical relaxation is then studied by analyzing the evolution of the Shannon entropy, occupation entropy, and the number of principal components in the time-dependent many-body basis set encompassing the multiconfigurational time-dependent Hartree method for bosons (MCTDHB)~\cite{alon,stre} (see below). Statistical relaxation is characterized by an increase in entropy and the vanishing of its fluctuations. MCTDHB can in principle \cite{alon,stre} and practice \cite{exact,exact2,exact3} provide exact solutions of the the time-dependent many-body Schr\"odinger equation (TDSE). Herein, we use the MCTDHB method for solving the TDSE at a high level of accuracy.
Let us emphasize that exploring the solutions of TDSE for systems of interacting particles in an external trap potential is a fundamental problem in many-body physics. A key aspect of the present study with the MCTDHB approach is to enable faithful tests of the properties of thermalization processes in quantum many-body systems by employing an exact and realistic set of many-body wavefunctions as a basis for the computation of entropies and other relevant quantities. Let us add that the present work is restricted to the many-body dynamics of a pure state, i.e. the temperature of the system is absolute zero. 

We define the many-body information entropy $S^{info}(t)$ and the number of principal components $N_{pc}(t)$ in terms of the time-dependent expansion coefficients of the state in the time-dependent MCTDHB basis. These measures are then compared with the occupation entropy $S^{occu}(t)$ which is defined by the eigenvalues of the reduced one-body density matrix or occupation numbers, see Sec.~\ref{Squantities} and Refs.~\cite{socc,socc1}. As a key result we demonstrate that $S^{info}(t)$ and $S^{occu}(t)$ behave in a similar way: for increasing interparticle interactions, the many-body entropies saturate with time and their fluctuations become negligible. Hence, there is statistical relaxation \cite{xxx,sant3}, despite the tendency of the MCTDHB description to adapt the basis to minimize the number of contributing coefficients. Importantly, we also demonstrate that $S^{info}(t)$ and $S^{occu}(t)$ approach the value predicted by the GOE random matrix ensemble \cite{vkota,haak,meht} for larger 
values of the interaction strength. 

To complement the results for the entropies $S^{info}(t)$ and $S^{occu}(t)$ we study the time-evolution of the correlation function \cite{RJG,RDMs} that quantifies the coherence and fringe visibility in interference experiments. For a Shannon entropy defined with a noninteracting basis set it was shown in
Ref.~\cite{berm} that the interference fringes in ballistic expansion become less visible in the case of irregular dynamics, i.e., when the entropies become large. We investigate the relation of the correlation function to the production of many-body entropy quantitatively: we compare the time-evolution of the spatial first-order correlation function or coherence $g^{(1)}$ and the many-body information entropy $S^{info}(t)$. Our results demonstrate the close
relation of the loss of coherence and increase in many-body entropy. Since the spatial
correlation functions can be measured \cite{corr1,corr2}, this relation can
be scrutinized in experiments to test statistical relaxation directly. Let us recall here that testing the ETH means to verify that statistical relaxation is present in the system's many-body observables \cite{horo,brod,kota1,sant2,sant3}. Measuring the expectation values of general many-body operators is however a difficult if not impossible task. Our results demonstrate that this problem can be circumvented, since statistical relaxation can be inferred from measuring the one-body correlation function.

This paper is organized as follows. In Sec.~II we give the Hamiltonian used
in the present work and a brief introduction to the numerical many-body method
MCTDHB as well as the quantities of interest. 
Sec.~III discusses the time-evolution of many-body entropy measures as a function of the interaction strength to which the system is quenched. Sec.~IV considers
correlation functions and coherence in the dynamics and demonstrates a link of statistical relaxation
to the loss of coherence. Sec.~V gives conclusions of our work.

\section{Methodology}

\subsection{Numerical method and Hamiltonian}
The evolution of $N$ interacting bosons is governed by the TDSE,
\begin{equation}
\hat{H} \Psi = i \frac{\partial \Psi}{\partial t}. \label{TDSE}
\end{equation}
The total Hamiltonian we consider is
\begin{equation}
\hat {H} (x_{1}, x_{2}, ..., x_{N}) = \sum _{i=1}^{N} \hat {h}(x_{i}) + 
\Theta(t) \sum _{i<j=1}^{N} \hat {W} (x_{i}-x_{j}). \label{Many-Body-H}
\end{equation}
Here, $\hat{h}(x) = \hat{T}(x) + \hat{V}(x)$ is the one-body Hamiltonian containing the
external trapping potential $\hat{V}$ and the kinetic energy
$\hat{T}=-\frac{1}{2}\hat{\partial}^2_x$, $\hat {W} (x_i-x_j)$ is the two-body
interaction of particles at positions $x_i,x_j$, and $\Theta(t)$ is the Heaviside step function of time $t$. The Hamiltonian $\hat{H}$ is in dimensionless units. It is obtained by dividing the dimensionful Hamiltonian by $\frac{\hbar^2}{mL^2}$ ($m$ is the mass of the bosons, $L$ is an arbitrary length scale). Since the time-evolution
starts at $t=0$, the $\Theta(t)$ term in Eq.~\eqref{Many-Body-H} above
implements an interaction quench: the interactions are abruptly turned on at
$t=0$. We initialize the system in the ground-state $\vert \Psi (t=0) \rangle$ of the noninteracting Hamiltonian (Eq.~\eqref{Many-Body-H} for $t<0$). In the MCTDHB method which we use to solve the TDSE, Eq.~\eqref{TDSE}, with the Hamiltonian,
Eq.~\eqref{Many-Body-H}, the ansatz for the many-body wave function is taken as a linear combination of time-dependent permanents with time-dependent weights,
\begin{equation}
\vert \psi(t) \rangle = \sum_{\vec{n}}C_{\vec{n}}(t) \vert \vec{n};t \rangle; 
\qquad \vert \vec{n};t\rangle = \prod_{i=1}^M \left[\frac{\left(
\hat{b}_i^\dagger(t)\right)^{n_i}}{\sqrt{n_i!}}  \right] \vert vac \rangle. 
\label{ansatz}
\end{equation}
Here, the summation runs over all possible $N_{conf}=\binom{N+M-1}{N}$ configurations, $\lbrace \vert \vec{n}; t \rangle = \vert n_1,...,n_M; t \rangle; \sum_i n_i \equiv N \rbrace$, $\hat{b}_i^\dagger(t)$ creates a boson in the $i$th
single-particle state $\phi_i(x,t)$, and $\vert vac \rangle$ denotes the vacuum.
It is important to emphasize that in the ansatz [Eq.~\eqref{ansatz}] both the
expansion coefficients $\lbrace C_{\vec{n}} (t); \sum_i n_i = N
\rbrace$ \textit{and} the orbitals $\lbrace \phi_i(x,t)
\rbrace_{i=1}^M$ that build up the permanents $\vert \vec{n}; t \rangle$ are
time-dependent, fully variationally optimized quantities. MCTDHB has been
established as the currently most efficient way to solve the time-dependent
many-body problem of interacting bosons accurately and for a wide set of
problems \cite{exact,exact2,exact3}. In MCTDHB(M), the vectors $\vec{n} = \left(n_{1},...,n_{M} \right)$ represent the occupations of the orbitals in a single
configuration and preserve the total number of particles, $n_{1}+...+n_{M} = N$.
$M$ is the number of single-particle functions that make up the permanents $\vert
\vec{n}; t \rangle$. The efficiency of the method comes from the variationally
optimized and time-adaptive basis that makes the sampled Hilbert space dynamically
follow the motion of the many-body dynamics. 

In the limit of $M \rightarrow \infty$ the set of permanents  $\lbrace \vert \vec{n};t \rangle \rbrace$ spans the complete $N$-boson Hilbert space and the expansion in
Eq.~\eqref{ansatz} is exact. In practice, we have to limit the size of the
Hilbert space in our computations. Because the permanents are time-dependent, a given degree of accuracy is reached with a much shorter expansion, as compared to a
time-independent basis. This leads to a significant computational advantage over, for instance, exact diagonalization techniques (see explanation
below). To solve the TDSE, Eq.~\eqref{TDSE}, for the wave function $\vert \Psi(t) \rangle$ one needs to
determine the evolution of the coefficients $\lbrace C_{\vec{n}} (t) \rbrace$ and orbitals $\lbrace \phi_i(x,t) \rbrace_{i=1}^M$ in time. Their equations of motion are derived by requiring the stationarity of the
action functional with respect to variations of the time-dependent
coefficients and the set of time-dependent orbitals. The obtained equations form
a coupled set of nonlinear integro-differential equations \cite{alon} that we solve simultaneously with the recursive MCTDHB (R-MCTDHB) package \cite{ultr}. For reference, we also give the equations of motion in Appendix~\ref{EOM}. In order to calculate
eigenstates of the Hamiltonian $\hat{H}$ [Eq.~\eqref{Many-Body-H}], one uses the
so-called improved relaxation method. By propagating the MCTDHB equations of motion [Eqs.~\eqref{OEOM} and \eqref{CEOM}] in imaginary time for a given initial guess, excitations are exponentially damped and the system relaxes to the ground state. It should be noted that in the widespread time-dependent Gross-Pitaevskii (TDGP)
theory, the many-body wave function is given by a single permanent $\vert n_1=N;
t \rangle$; all particles reside in the same single-particle state (orbital) and there is consequently
only a single coefficient. MCTDHB($M$) contains the TDGP theory as the $M = 1$ special
case.

We stress here that MCTDHB is much more accurate
than exact diagonalization methods at the same dimensionality of the considered
space. In exact diagonalization a time-independent basis is employed. In most cases, it is built from
the eigenstates of a one-body problem. These states are not further optimized to
take into account the dynamics and correlations in the considered system which necessarily
arise due to the presence of inter-particle interactions. Thus the space and basis
considered in exact diagonalization is \textit{fixed} and \textit{not optimized}, especially for
the treatment of dynamics. MCTDHB on the other hand uses a time-adaptive
many-body basis set (see Eq.~\eqref{ansatz}, Ref.~\cite{alon}, and also Appendix
\ref{EOM}). Its evolution follows from the time-dependent variational principle \cite{TDVP} and is such that the error resulting from the truncation of many-body Hilbert space
\textit{is minimized by the basis at any given point in time}. This advantage of
time-adaptivity helped to solve \textit{numerically exactly} the time-dependent
many-body problem even for long-range time-dependent one-body and two-body potentials, see
Refs.~\cite{exact,exact2}. These references show in a direct comparison of
MCTDHB and exact diagonalization that the accuracy of MCTDHB for many-body \textit{dynamics} of
ultracold bosons can in many cases not be reached at all by exact diagonalization methods, even when a very
large configuration space is used. 

The Hamiltonian of bosons in one dimension is given by the
Lieb-Liniger model when the two-body interaction is assumed to be mediated by a
contact potential \cite{lieb}. In the limit of $\frac{n}{g}
\rightarrow 0$ (here $n$ is the particle density, $g$ is the inter-atomic
coupling strength) fermionization occurs. This so-called Tonks-Girardeau 
regime is characterized by a single but strongly correlated configuration~\cite{Gir}, i.e., the expansion Eq.~\eqref{ansatz} would include only a single term. On the other extreme, the $\frac{n}{g}\rightarrow \infty$ limit can be captured by the TDGP mean-field approximation for weakly interacting bosons, i.e., a single uncorrelated configuration. One could hence speculate that an entropy measure related to
configuration space would decrease as the system enters the Tonks-Girardeau regime since the eigenstates are single configurations.
However, in our study we did not encounter such a behavior. This means that
the dynamics in our study is far from equilibrium and adiabaticity and the Tonks-Girardeau states do not play much of a role. Moreover, we chose interaction strengths, for which $\frac{n}{g}\rightarrow \infty$ is not fulfilled sufficiently well; our investigations are therefore in the crossover between the two regions $\frac{n}{g}\rightarrow \infty$ and $\frac{n}{g}\rightarrow 0$ where any mean-field approach (Gross-Pitaevskii \cite{GPbook} or multi-orbital \cite{TDMF}) breaks down, because many different configurations are contributing. In this regime, as we will see in the following, many-body entropies and correlations become important and their features are present in the quantum dynamics.

\subsection{Quantities of interest}
\subsubsection{Many-body entropies}\label{Squantities}
To study statistical relaxation and thermalization, we employ the measures information entropy $S^{info}(t)$ and number of principal components $N_{pc}(t)$ defined as follows 
\begin{eqnarray}
N_{pc}(t) &=& \frac{1}{ \sum_{\vec{n}} |C_{\vec{n}}(t)|^{4} }, \\
S^{info}(t) &=& - \sum_{\vec{n}}  |C_{\vec{n}}(t)|^{2} 
\ln |C_{\vec{n}}(t)|^{2}. \label{MBentr}
\end{eqnarray}
$S^{info}(t)$ and $N_{pc}(t)$ measure the effective number of basis states that
contribute to a given many-body state at time $t$. The many-body nature of these quantities for the MCTDHB basis set [Eq.~\eqref{ansatz}] can be made explicit by writing a coefficient $|C_{\vec{n}}(t)|^2$ as an expectation value,
\begin{equation*}
 |C_{\vec{n}}(t)|^2 = \frac{1}{\prod_{i=1}^M n_i !} \langle \Psi \vert 
 (\hat{b}_1(t))^{n_1} \cdots (\hat{b}_M(t))^{n_M}  (\hat{b}^\dagger_1(t))^{n_1}
 \cdots (\hat{b}^\dagger_M(t))^{n_M} \vert \Psi \rangle \;.
\end{equation*}
Obviously, all $M$ creation/annihilation operators contribute to the value of
every coefficient. For this reason, the coefficients and their distribution can also be used to directly assess the content of many-body entropies in the the system qualitatively.

In the multi-orbital \cite{TDMF} and Gross-Pitaevskii
\cite{GPbook} mean-field approaches only a single configuration and
coefficient is included and consequently $S^{info}(t)\equiv 0$ and $N_{pc}(t) \equiv 1$.
Thus, $S^{info}(t)$ and $N_{pc}(t)$ are entropies that cannot be produced in
mean-field theories. The information entropy and number of principal components of the MCTDHB basis therefore are a quantitative measure for how well or not a given many-body state is captured by mean-field theories. Large $N_{pc}(t)$ or $S^{info}(t)$ means a state contains many configurations and cannot be captured by mean-field methods. Small $N_{pc}(t)$ or $S^{info}(t)$ means that the state contains few configurations and is close to a mean-field state.

Besides the two measures $S^{info}(t)$ and $N_{pc}(t)$, we consider also the occupation entropy \cite{socc,socc1} defined by 
\begin{equation}
S^{occu}(t) = -\sum_{i} {\bar{n}}_{i}(t) \Big[ \ln \hspace*{.1cm} {\bar{n}}_{i}(t) \Big] .\label{Socc}
\end{equation}
$S^{occu}(t)$ is an entropy obtained from the natural occupations, i.e., the eigenvalues of the reduced one-body density matrix $\bar{n}_{i}(t)=\frac{n_i(t)}{N}$ (see Eq.~\eqref{eq.13} below and Ref.~\cite{RDMs} for details). 
For the TDGP mean-field one has $S^{occu}(t)=0$ always, since there is only one natural occupation $\bar{n}_1= n_1/N=1$ in this case. For multi-orbital mean-field theories, several occupation numbers can be different from $0$; however, these occupations are time-independent, i.e., $\partial_t \bar{n}_i = 0$. Hence, for multi-orbital mean-field theories the occupation entropy $S^{occu}(t)$ remains constant, i.e., $\partial_t S^{occu}(t)=0$.
In the time-evolution of a many-body state, the proximity of the value of $S^{occu}(t)$ to $0$ is a measure of how well the state can be described by the TDGP mean-field. The magnitude of the fluctuations in $S^{occu}(t)$ indicates how well the state could be described by a multi-orbital mean-field approach. Since there are non-mean-field states with many configurations that also fulfill $\partial_t S^{occu}(t)=0$, it has to be checked if additionally $S^{info}(t)\approx0$ holds in order to conclude that a state with constant occupation entropy is indeed of single-configuration (i.e., mean-field) type.

For a GOE of random matrices, $N^{GOE}_{pc}=D/3$ and $S^{info}_{GOE}=\ln0.48D$ holds, where $D\times D$ is the dimension of the random matrices. To obtain the GOE estimate for the information entropy $S^{info}(t)$, we set $D$ equal to the number $N_{conf}$ of time-dependent many-body states in MCTDHB. For the occupation entropy, the GOE analog is obtained by setting $n_{i} = \frac{N}{M}$ for all $i$ in Eq.~\eqref{Socc}. Consequently, in the case of $S^{occu}(t)$, $D$ is set equal to the number of orbitals $M$ in the MCTDHB treatment. We obtain $S^{occu}_{GOE} = -\sum_{i=1}^{M}\left(\frac{1}{M}\right) \ln \left(\frac{1}{M}\right) = \ln(M)$ in this case.

\subsubsection{Correlation functions and coherence}\label{gquantities}
In the present work, we investigate the normalized first-order correlation function 
$g^{(1)}(x_{1}^{\prime},x_{1},t)$ defined as \cite{RDMs,RJG} 
\begin{equation}
g^{(1)}(x_{1}^{\prime},x_{1};t) = \frac{ \rho^{(1)}(x_{1}|x_{1}^{\prime};t) }{
\sqrt{\rho(x_{1},t)\rho(x_{1}^{\prime},t)}}
\end{equation}
where $\rho$ is the diagonal part of the one-body density matrix $\rho^{(1)}$ given by
\begin{eqnarray}
\rho^{(1)} (x_{1}|x_{1}^{\prime};t)  = \hspace*{5cm}& \hspace*{0.6cm}  
\nonumber \\ 
N \int \psi^{*}(x_{1}^{\prime},x_{2},\dots,x_{N};t)  \psi(x_{1},x_{2},\dots,x_{N};t) 
dx_{2} dx_{3}\dots dx_{N}. &
\label{eq.13}
\end{eqnarray}

The normalized spatial first order correlation function $g^{(1)}$ quantifies the
degree of first order coherence. Note that $|g^{(1)}(x_{1}^{\prime},x_{1};t)| < 1$ means that the visibility of interference fringes in interference experiments will be less than $100\%$; this case is referred to as \textit{loss of coherence}.
$|g^{(1)}(x_{1}^{\prime},x_{1};t)| = 1$ corresponds to maximal fringe visibility
of the interference pattern in interference experiments and is referred to as \textit{full coherence}. The strength of the interaction between the particles affects the correlations: the stronger the inter-particle repulsion, the stronger the loss of coherence.

\section{The production of many-body entropy in interaction quench 
dynamics}\label{Results}

Our present calculations are performed for one spatial dimension and consider $N=10$
repulsively interacting bosons in an external harmonic trap. We use a contact
interaction $\hat{W}(x_{i}-x_{j})=\lambda_0 \delta(x_{i}-x_{j})$ and the
external trap $\hat{V}(x_{i}) = \frac{1}{2} x_{i}^{2}$ in
Eq.~\eqref{Many-Body-H}. The dimensionless strength $\lambda_{0}$ of the
inter-particle repulsion is varied in our investigation. 
We restrict the number of orbitals to $M=6$, yielding a total of
$D\equiv N_{conf}=\binom{N+M-1}{N}=3003$ permanents. A key aspect of many-body quantum
chaos is how the many-body wave function spreads across the available basis
functions with an increase in the inter-particle interaction strength. As mentioned,
this spreading is counteracted by the time-adaptivity of the basis set in
MCTDHB.

For an assessment of the presence of statistical relaxation, it is
instructive to first visualize the time-evolution of the coefficients
$C_{\vec{n}}(t)$ directly. In Figs.~1 and 2 we plot $|C_{\vec{n}}(t)|^{2}$ as a function
of the index $n$ of the basis states $\vert \vec{n};t \rangle$.
See Ref.~\cite{MAP} for the details on how to obtain the index $n$ from the vectors $\vec{n}=(n_1,...,n_M)$. Fig.~1 is for the interaction strength $\lambda_{0} = 0.5$ and for times $t = 0.2, 0.5, 1.0$, and $2.0$. Note that the permanents $\vert
\vec{n};t \rangle$ are not eigenstates of the Hamiltonian for $t > 0$. However,
every eigenstate of the interacting problem can be represented as a pattern of
coefficients that contribute significantly. Initially, at $t=0$, only a single 
coefficient is nonzero, because the system was relaxed to the noninteracting
ground state. As depicted in Figs.~1a)-d), the number of significantly contributing coefficients
grows with time due to the inter-particle interactions, but remains only a small portion of the available basis states for $M=6$.
This is a consequence of the interactions being relatively weak. If the number of
non-zero elements of $\lbrace C_{\vec{n}} (t) \vert \sum_i n_i \equiv N \rbrace$ is only
a small portion of $N_{conf} = 3003$, we refer to the respective state as \textit{localized}. 
Localized states are rather close to a mean-field description for which only a single coefficient would
contribute. The results in Fig.~2 are for a larger value of the interaction strength
($\lambda_{0} = 10.0$). It is seen from Fig.~2 that there is a substantially
larger amount of non-zero coefficients as compared to the smaller
interaction strength $\lambda_0=0.5$ (Fig.~1). Since the contributing coefficients
spread over almost the whole available space, we refer to such a state as
\textit{delocalized}. Delocalized states cannot be captured by mean-field descriptions. From Figs.~1 and 2 we conclude that increasing the interaction strength makes the states emerging in the time-evolution change character from localized to delocalized.

To further quantify the average number of time-dependent basis states that make up the
wave function, we plot $\exp(S^{info}(t))$ and $N_{pc}(t)$ for the small interaction strength $\lambda_{0} = 0.5$ in Fig.~3, and for the large interaction strength $\lambda_{0} = 10.0$ in Fig.~4. For $\lambda_0=0.5$, the inter-atomic correlations are small and large fluctuations in both $N_{pc}(t)$ and $\exp(S^{info}(t))$ are seen (Fig.~3). These fluctuations are around comparatively small averages (compare Figs.~3 and 4). We conclude that for localized states, there are only a few contributing components. For larger interaction strength (Fig.~4), the fluctuations in $\exp(S^{info}(t))$ and $N_{pc}(t)$ decrease. This is analog to the GOE behavior and  in line with computations employing a time-independent basis set \cite{bohi, haak, meht}. An important observation from Fig.~4 is that the entropies saturate for large interaction strength. We can conclude from Figs.~3 and 4 that increasing $\lambda_0$ makes the fluctuations of the number of principal components $N_{pc}(t)$ and $\exp(S^{
info}(t))$ decrease and results in the emergence of a saturation of these quantities.

To shed more light on this behavior, we continue by comparing the many-body entropies $S^{info}(t)$ and $S^{occu}(t)$. 

Fig.~5 shows $S^{info}(t)$ and $S^{occu}(t)$ (Eq.~\eqref{MBentr} and \eqref{Socc}, respectively) for the interaction strengths $\lambda_0=0.5, 10.0$, and $15.0$. As a first observation, we find that $S^{info}(t)$ and $S^{occu}(t)$
show a similar overall behavior. We discuss first the information entropy $S^{info}(t)$ in detail and subsequently $S^{occu}(t)$. 

For smaller interactions, $\lambda_0 = 0.5$, the increase in $S^{info}(t)$ is almost linear and reaches the value $S^{info}\simeq 3.2$ at time $t=6$ [Fig.~5a)]. The GOE value $S^{info}_{GOE}$ for our present calculation
with $10$ bosons, $6$ orbitals, and $D=N_{conf} = 3003$ is $S^{info}_{GOE} = ln(0.48D)
= 7.273$. Since we observe $S^{info}(t)<S^{info}_{GOE}$, we infer that the system remains relatively ordered. Nevertheless, the value of $S^{info}(t)$ is far enough from $0$ to conclude that the state can no longer appropriately be described by mean-field methods. For larger interaction strengths, $S^{info}(t)$ shows quick saturation [Fig.~5b) and c)]. For $\lambda_{0} = 10.0$, saturation emerges close to $S^{info}=6.49$ and for $\lambda_{0} = 15.0$ at $S^{info}=7.17$. The values of $S^{info}(t)$ are approaching the GOE value $S^{info}_{GOE}=7.23$. The discrepancy is because the interaction strength, albeit being comparatively large, is still finite. 
As far as time-independent basis sets are concerned this may be attributed to the operation of the so-called embedded GOE of random matrices in interacting particle systems (Ref.~\cite{sahu}). 
The theory of embedded GOEs can be applied to systems with lower-body rank operators such as the one- plus two-body Hamiltonian used in the present study. It predicts that entropy measures will be close to but not identical to the GOE value \cite{vkota}. Importantly, the saturation of the many-body information entropy $S^{info}(t)$ close to the GOE value $S^{info}_{GOE}$ can be seen as a hallmark of statistical relaxation. This saturation demonstrates that statistical relaxation \textit{gradually overcomes} the tendency of the time-dependent variational principle to minimize the spread of the coefficients by optimization of the MCTDHB basis set as the inter-particle
repulsion $\lambda_0$ increases. Let us now consider the occupation entropy $S^{occu}(t)$ in more
detail.

In Fig.~5, the results for $S^{occu}(t)$ are shown for the interaction strengths $\lambda_0=0.5,10.0$, and
$15.0$. At $t=0$, all bosons are in the lowest orbital: $n_{1}=N$ and $n_{i}=0$ for $i = 2,3...M$. Then $\bar{n}_{1}=1$ and all other $\bar{n}_i=0$ [cf. Eq.~\eqref{Socc}]. Therefore $S^{occu}(t=0) = -\ln(1) = 0$ as in the case of the information entropy $S^{info}(t=0)$.
The system is, hence, in a mean-field state at $t=0$ (cf. Sec.~\ref{Squantities}). At later times, however, the bosons start to distribute themselves in all $M$ orbitals. 
For all interactions [Figs.~5a),b), and c)] $S^{occu}(t)$ is saturating after an initial adjustment to the quench. The state is, however, not of (multi-orbital) mean-field character, since $S^{info}(t)>0$ indicates that several configurations are contributing to the dynamics. 
For small interactions [Fig.~5a)], $S^{occu}(t)$ saturates far below the value of the GOE. The state remains rather localized and the dynamics can be considered as regular and we refer to the state as \textit{ordered}.
For larger values of the interaction strength ($\lambda_0=10.0$ and $\lambda_0=15.0$ in Figs.~5b) and 5c), respectively), when the wave function becomes fully delocalized with time, all bosons are on
average roughly equally distributed in the $M$ orbitals. The situation resembles the GOE and $S^{occu}(t)$ saturates close to its GOE estimate $S^{occu}_{GOE}$. In our calculations with $M=6$ and $N=10$, $S^{occu}_{GOE} = \ln (M) = \ln(6) = 1.79$, as discussed in Sec.~\ref{Squantities}. This convergence of $S^{occu}(t)$ to $S^{occu}_{GOE}$ for increasing $\lambda_0$ demonstrates the presence of statistical relaxation and is similar to the many-body information entropy $S^{info}(t)$ discussed in the previous paragraph. As also argued for $S^{info}(t)$, the value at which $S^{occu}(t)$ saturates is smaller than the GOE value $S^{occu}_{GOE}$ due to the applicability of the theory on embedded GOEs for large but finite interactions. One may speculate that the chaotic or irregular dynamics which we find are a many-body analog of the wave chaos found for the TDGP case in Ref.~\cite{socc1}.

Fig.~6 shows the information entropy $S^{info}(t)$ for much longer times than Fig.~5. For interaction strength $\lambda_0=0.5$, $S^{info}(t)$ does not reach saturation and
exhibits strong fluctuations for a long time [Fig.~6a)]. The system may eventually reach some equilibrium
state which is different from a thermal state and may, according to Ref.~\cite{Rigol}, be described in terms of a
generalized Gibbs ensemble. For $\lambda_0=15$ it is seen from Fig.~6b) that there is a plateau at small times ($t \sim 0.5$, see Fig.~6b) inset). The relaxation to the GOE value $S^{info}_{GOE}$ happens on a much longer time scale ($t \sim 10$) and is non-transient. We speculate that the first stage of relaxation on a much shorter time scale may be a signature of so-called \textit{prethermalization}. Prethermalization is characterized by a rapid relaxation of \textit{only some} observables (not all, as in the case of the ETH) to their equilibrium values~\cite{Berges} and has recently been observed experimentally for a degenerate one dimensional Bose gas~\cite{Langen}.

\section{Relation of correlation function and coherence to many-body entropy production}

In this section we discuss the time-evolution of
the first-order correlation function $g^{(1)}$ (see Sec.~\ref{gquantities} for a definition) for our system. 
The correlation function quantifies the coherence which can be experimentally determined in interference experiments. We would like to put forward a strong link between the dynamics in the coherence and in the many-body measures of entropy defined in Sec.~\ref{Squantities} and analyzed in the previous Section. This connection can be used to quantify the many-body entropy in the system by measuring the correlation function.

Fig.~7 presents $|g^{(1)}(x_{1}^{\prime},x_{1};t)|^2$ as a function of its two spatial
variables for two interaction strengths, $\lambda_0=0.5$ and $\lambda_0=10$ at various times $t$. For
weak interactions, i.e., $\lambda_0=0.5$,  $|g^{(1)}(x_{1}^{\prime},x_{1};t)|^2$ remains close to unity for all $(x_{1}^{\prime},x_1)$ for a comparatively long time. The system remains coherent throughout its time-evolution. This observation is in-sync with the small spreading of the states coefficient distribution (see Fig.~1) for small interactions $\lambda_0=0.5$. Turning to stronger interaction strength
($\lambda_0=10.0$), after a sufficiently long time  ($t \sim 10$) the correlation
function is unity almost only along the diagonal ($x_{1}^{\prime}=x_1$). Away
from the diagonal ($x_1^{\prime}\neq x_1$) the correlation function $|g^{(1)}(x_{1}^{\prime},x_{1};t)|^2$ is close to $0$. Hence, for the stronger interactions $\lambda_0=10.0$, the coherence of the system is lost with time.
That strong inter-particle repulsion leads to an almost complete \textit{loss of coherence} agrees with our previous observation of statistical relaxation, saturation of entropies $S^{info}(t)\rightarrow S^{info}_{GOE}$ and $S^{occu}(t)\rightarrow S^{occu}_{GOE}$, and delocalization of the MCTDHB coefficients distribution (compare Figs.~1,5, and 7). We conclude that the production of many-body information entropy $S^{info}(t)$ and occupation entropy $S^{occu}(t)$ entails an intensified loss of coherence. This connection can be exploited to measure the many-body entropy of a system, because the one-body correlation function $g^{(1)}$ can be determined through interference experiments. 
Let us stress again here that the larger the disorder [measured by the entropies $S^{info}(t)$ and $S^{occu}(t)$] in a quantum mechanical system the less it can be described by a product of a single complex valued function. Since the ansatz of the TDGP uses a single complex-valued function, our finding implies that the TDGP can generally not adequately describe processes such as statistical relaxation or thermalization which are characterized by a loss of coherence and an in-sync increase and saturation of the entropies $S^{info}$ and $S^{occu}$.

\section{Conclusions}

We have studied many-body entropy production, statistical relaxation and
coherence of parabolically trapped interacting bosons for an interaction quench by highly accurate MCTDHB computations going beyond the scope of commonly applied mean-field approaches. The full
time-dependent solution of the many-body problem with MCTDHB allowed us to define and
compute new many-body entropy measures. We analyzed the information entropy, number of
principal components, and occupation entropy of the time-dependent MCTDHB basis set as a function of time. 
We have shown that an increase in the many-body entropy measures is linked to a loss of coherence in the dynamics (see Figs.~5-7). For larger values of the interaction strength, we find irregular dynamics and statistical relaxation despite the tendency of the time-adaptive many-body basis set of
MCTDHB to minimize the number of contributing expansion
coefficients and therewith the entropies related to them. All the entropy measures are in
mutual agreement (see Figs.~3-6).

For larger inter-particle repulsion, the expansion coefficients $|C_{\vec{n}}(t)|^2$ delocalize more strongly (Fig.~2) and we
observe a quick initial saturation of many-body entropies (see Figs.~5 and 6) possibly related to prethermalization. A saturation of the many-body entropy to the GOE value follows on a longer time scale. It is important to stress here that this saturation is not as trivial as one might be tempted to
assume: the basis set in MCTDHB is explicitly time-dependent and it is optimized such that it minimizes the portion of significant coefficients in the expansion. Nevertheless, the many-body entropy of the coefficients \textit{and} the eigenvalues of the reduced one-body density matrix approach the GOE values and
statistical relaxation prevails. 

By studying the time-dependence of the first order correlation we demonstrate a
strong link between the dynamics of entropy and the dynamics of coherence. Our
present work exemplifies that large production in many-body entropy causes an intensified loss of coherence (Figs.~5-7). This loss of coherence constitutes an independent signature of the existence of statistical relaxation, allowing to study it from another perspective and, most importantly, to measure
it in experiments. 

Further investigations are needed to test the ETH and to
assess the generality of statistical relaxation as well as its relation to chaos,
also for larger portions of many-body Hilbert space. In this respect it is
interesting to understand if the found emergence of statistical relaxation is a
many-body analog of the wave chaos in the TDGP equation reported in Ref.~\cite{socc1}. Finally, a straightforward continuation of this work would be to find its connection to investigate the possible connection the recently observed prethermalization phenomenon \cite{Langen}. 

\begin{acknowledgments}
We would like to thank Sudip Haldar for pointing out the connection of our work to the prethermalization phenomenon and Marios C. Tsatsos for helpful comments on the manuscript.

The hospitality of the Theoretical Chemistry Group and especially Lorenz S.
Cederbaum in Heidelberg, as well as helpful discussions with Ofir E. Alon are
gratefully acknowledged. Axel U.J. Lode acknowledges financial support by the
Swiss SNF and the NCCR Quantum Science and Technology. Barnali Chakrabarti
wishes to thank the Theoretical Chemistry Group of the University of Heidelberg
for providing financial assistance for her visit to the group of Lorenz S.
Cederbaum, where the work was started and acknowledges financial support of the
Department of Science and Technology, Govt. of India, under the major research
project [SR/S2/CMP-126/2012]. 

\end{acknowledgments}

\appendix

\section{The MCTDHB equations of motion}\label{EOM}

In the following, the equations of motion of MCTDHB are given and their
derivation is sketched. For details, see Ref.~\cite{alon}. The action of the TDSE, 
\begin{equation}
\mathcal{S} = \int dt \left( \langle \Psi \vert \hat{H}- i\partial_t \vert 
\Psi \rangle+ \sum_{ij} \mu_{ij}(t)\left( \langle \phi_i \vert \phi_j 
\rangle - \delta_{ij} \right) \right), 
\end{equation}
is demanded to be stationary with respect to the variation of the time-dependent
orbitals $\lbrace \phi_k(x;t) \rbrace_{k=1}^M$ \textit{as well as} with respect
to the variation of the time-dependent coefficients $\lbrace C_{\vec{n}}(t)\vert
\sum_i n_i =N\rbrace$. The orthonormality of the orbitals $\lbrace \phi_k(x;t)
\rbrace_{k=1}^M$ is enforced by the Lagrange multipliers $\mu_{ij}(t)$ in
$\mathcal{S}$. From the stationarity of the action $\mathcal{S}$ the equations
of motion of the orbitals,
\begin{eqnarray}
 i \partial_t \phi_j(x,t) &=& \mathbf{\hat{P}} \left. \Bigg(  \hat{h} 
\phi_j(x,t) + \sum_{k,s,q,l=1}^M \lbrace \rho(t) \rbrace^{-1}_{jk}  
\rho_{ksql}(t) \hat{W}_{sl}(x,t) \phi_q(x,t) \right. \Bigg), \label{OEOM} \\  
 \mathbf{\hat{P}} &=& \mathbf{1} - \sum_{j'=1}^M \vert \phi_{j'} \rangle 
 \langle \phi_{j'} \vert,\nonumber
\end{eqnarray}
as well as of the coefficients,
\begin{equation} 
 \mathcal{H}(t) \mathcal{C}(t) = i \partial_t \mathcal{C}(t);\qquad 
 \mathcal{H}_{\vec{m}\vec{m}'}(t) = \langle \vec{m};t \vert \hat{H} - 
 i \partial_t \vert \vec{m}' ; t \rangle, \label{CEOM}
\end{equation}
are derived. These equations \eqref{OEOM},\eqref{CEOM} form the core of MCTDHB.
Without loss of generality, equations \eqref{OEOM},\eqref{CEOM} are given here for one spatial dimension under the constraint $\langle \partial_t \phi_k \vert \phi_j \rangle = 0 \; \forall k,j$,
compare Ref.~\cite{alon}. The following notations were invoked for
the respective matrix elements and operators:
\begin{eqnarray}
 \rho_{kq}= \langle \Psi \vert \hat{b}^\dagger_k \hat{b}_q \vert \Psi \rangle, 
 \nonumber \\ 
 \rho_{ksql}= \langle \Psi \vert \hat{b}^\dagger_k \hat{b}^\dagger_s \hat{b}_q 
 \hat{b}_l\vert \Psi \rangle,\nonumber \\
 W_{sl}(x;t) = \int dx' \phi^*_s(x,t) W(x,x';t) \phi_l(x',t),\nonumber \\ 
 \hat{h}= -\frac{1}{2} \frac{\partial^{2}}{\partial x^{2}}+V(x)\nonumber.
\end{eqnarray}
Finally, the shorthand notation $\mathcal{C}(t)$ collects all the time-dependent
coefficients $\lbrace C_{\vec{n}}(t); \vec{n} \vert \sum_i n_i =N \rbrace$ in a vector,
employing an enumeration scheme documented in Ref.~\cite{MAP}.  The MCTDHB
equations of motion form a coupled set of nonlinear integro-differential
equations because the evaluation of the matrix elements $\rho_{kq},\rho_{ksql}$
in the orbitals equations \eqref{OEOM} depends on the coefficients
$\mathcal{C}(t)$, and the evaluation of the coefficients equation \eqref{CEOM}
depends on the matrix elements of the Hamiltonian with the current set of
orbitals, $h_{kq}=\langle \phi_k \vert \hat{h} \vert \phi_q \rangle$ and
$W_{ksql}=\int dx \int dx' \phi_k(x,t) \phi^*_s(x,t) W(x,x';t) \phi_l(x',t) \phi_q(x,t)$. The
equations \eqref{OEOM},\eqref{CEOM} can be solved efficiently and
self-consistently with the R-MCTDHB package \cite{ultr}.  For further details on
the derivation and properties of these equations, see Ref.~\cite{alon}.

\clearpage

\begin{figure}
  \begin{center}
     \includegraphics[width=\textwidth,angle=-90]{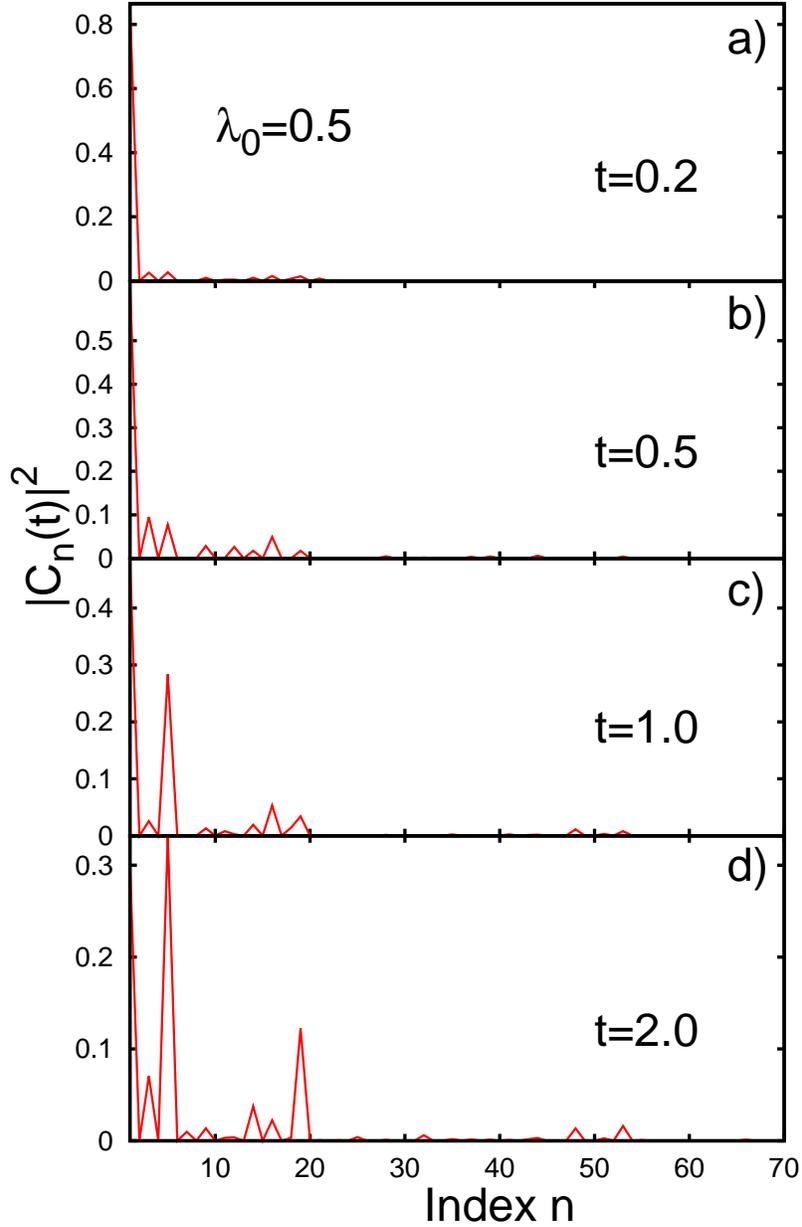}
  \end{center}

\caption{Time-evolution of the distribution of the magnitude of the
coefficients $\lbrace \vert C_{\vec{n}}(t) \vert^2 \rbrace$ for a weak
interaction, $\lambda_0=0.5$. Panels a)--d) show the magnitude of
the coefficients at times $t=0.2,0.5,1.0,2.0$, respectively. The index $n$ is
computed from the vector $\vec{n}$ using the mapping described in
Ref.~\cite{MAP}. With increasing time, more coefficients in the expansion become
significant, but the spread is far from the whole available space spanned by the
$D=N_{conf}=3003$ configurations. The state stays rather \textit{localized}
throughout the quench dynamics. The coefficients with $n>70$ are smaller than
$10^{-4}$ and not plotted therefore. All quantities are dimensionless. }

\end{figure}

\begin{figure}
  \begin{center}
       \includegraphics[width=\textwidth,angle=-90]{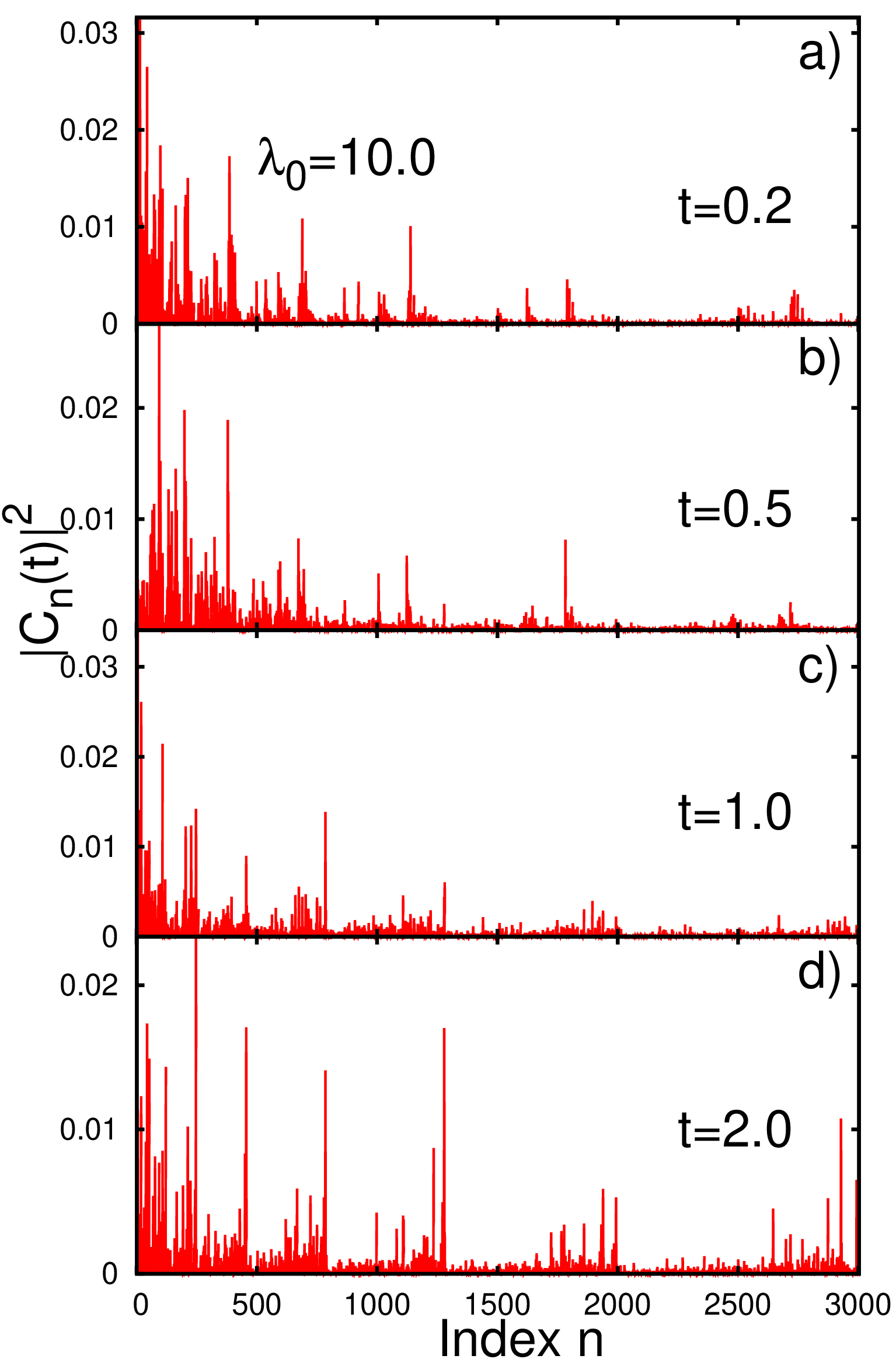}
  \end{center}

\caption{Time-evolution of the distribution of the magnitude of the coefficients
$\lbrace \vert C_{\vec{n}}(t) \vert^2 \rbrace$ for a strong interaction, 
$\lambda_0=10.0$. Panels a)--d) show the magnitude of the
coefficients at times $t=0.2,0.5,1.0,2.0$, respectively. The index $n$ is
computed from the vector $\vec{n}$ using the mapping described in
Ref.~\cite{MAP}. With increasing time, more coefficients in the expansion become
significant; the coefficients explore almost the whole available space spanned by
the $D=N_{conf}=3003$ configurations. The state rapidly becomes rather
\textit{delocalized} throughout the quench dynamics as compared to the localized
case in Fig.~1. All quantities are dimensionless.}

\end{figure}

\begin{figure}
  \begin{center}
       \includegraphics[width=10cm,angle=-90]{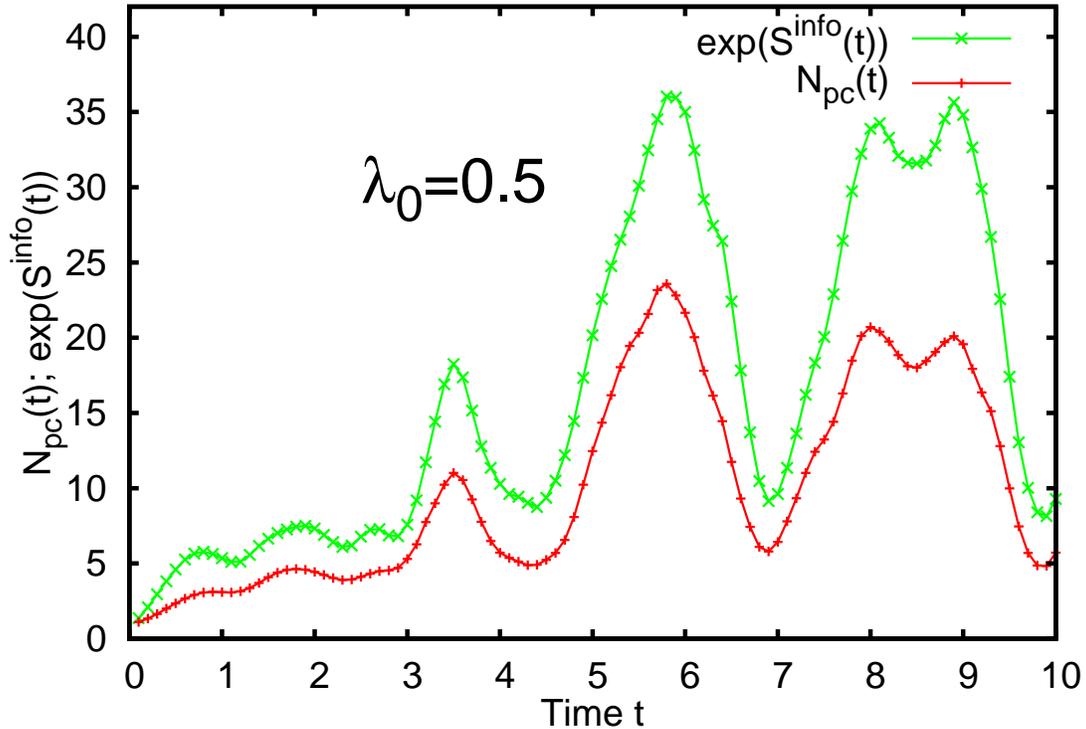}
  \end{center}

\caption{Time-evolution of the number of principal components $N_{pc}(t)$ and the exponential of information entropy $\exp(S^{info}(t))$ for the weak interaction $\lambda_0=0.5$. Both $\exp(S^{info}(t))$ (green, upper line) and $N_{pc}(t)$ (red, lower line) exhibit the same overall behavior. In analogy to the GOE for
small interactions, large fluctuations emerge in both quantities due to the
absence of strong correlations between the particles. See text for further discussion. All quantities
are dimensionless.}

\end{figure}

\begin{figure}
  \begin{center}
       \includegraphics[width=10cm,angle=-90]{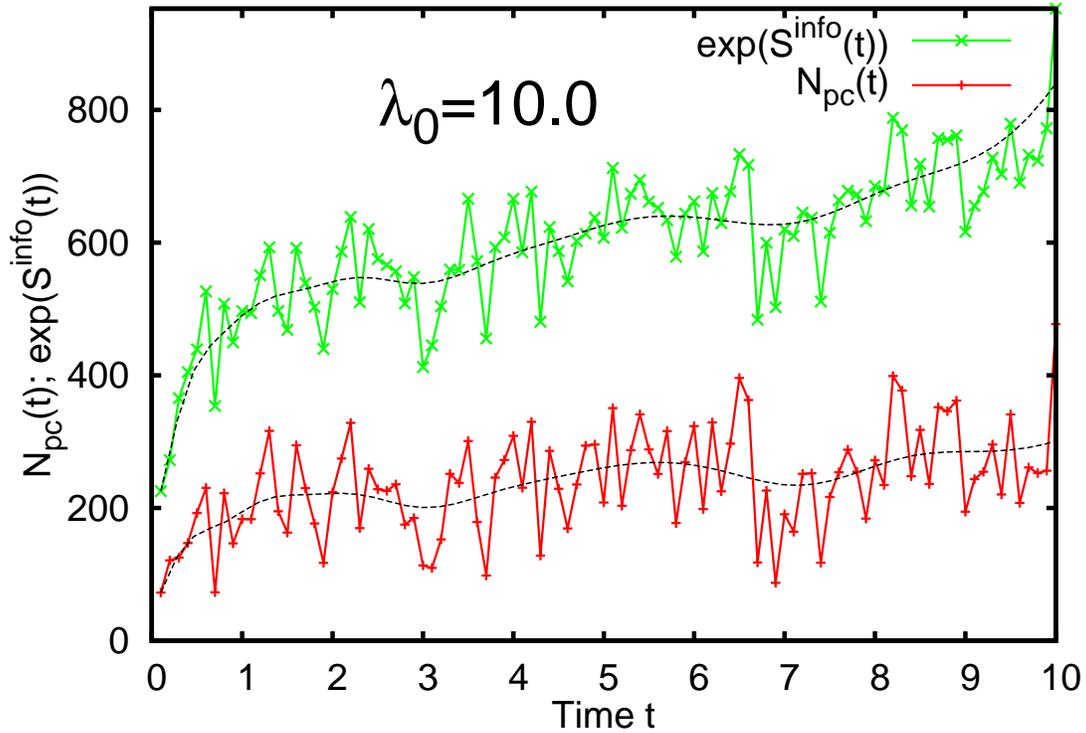}
  \end{center}

\caption{Time-evolution of the number of principal components $N_{pc}(t)$ and
the exponential of information entropy $\exp(S^{info}(t))$ for stronger interactions, $\lambda_0=10.0$. 
As in Fig.~3 for weaker interactions, both $\exp(S^{info}(t))$ (green, upper line) and $N_{pc}(t)$
(red, lower line) exhibit the same overall behavior. The thin black dashed lines
are provided to guide the eye and to estimate the magnitude of fluctuations. In analogy to the GOE for larger interactions, the relative fluctuations are quenched due to the
presence of strong correlations as compared to the case of smaller interactions
values in Fig.~3. See text for further discussion. All quantities are dimensionless.}

\end{figure}

\begin{figure}
  \begin{center}
  \includegraphics[width=10cm,angle=-90]{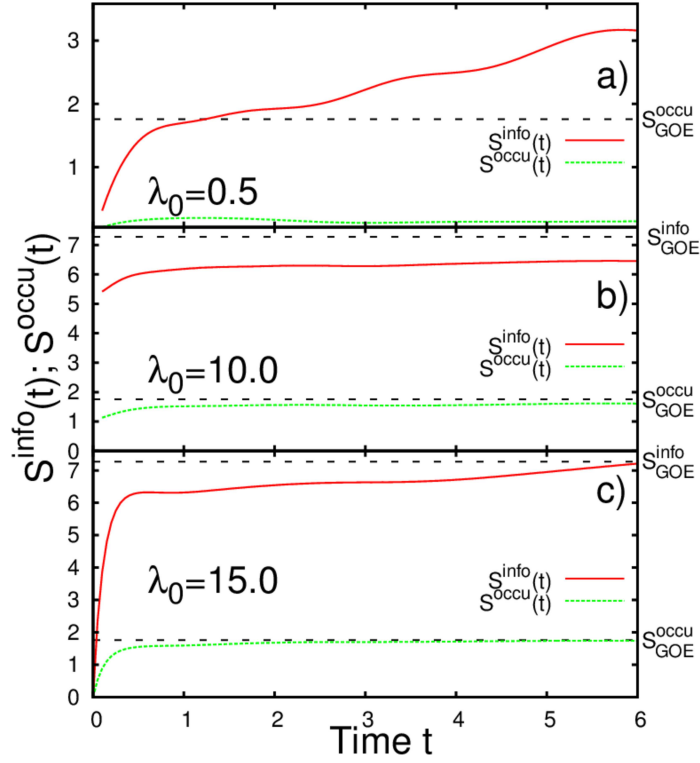}
  \end{center}

\caption{Dynamics of many-body Shannon entropy $S^{info}(t)$ and occupation
entropy $S^{occu}(t)$ for different inter-particle interaction strengths.
Statistical relaxation causes the convergence of both quantities,
$S^{info}(t)$ (red, upper lines) and $S^{occu}(t)$ (green, lower lines), to the
values $S^{info}_{GOE}$ and $S^{occu}_{GOE}$ (horizontal black dashed
lines), respectively, as the interaction strength increases from panels a) to
c). Implications are discussed in the text, curves are smoothened for clarity of presentation, all quantities shown are dimensionless.}

\end{figure}

\begin{figure}
  \begin{center}
    \includegraphics[width=10cm,angle=-90]{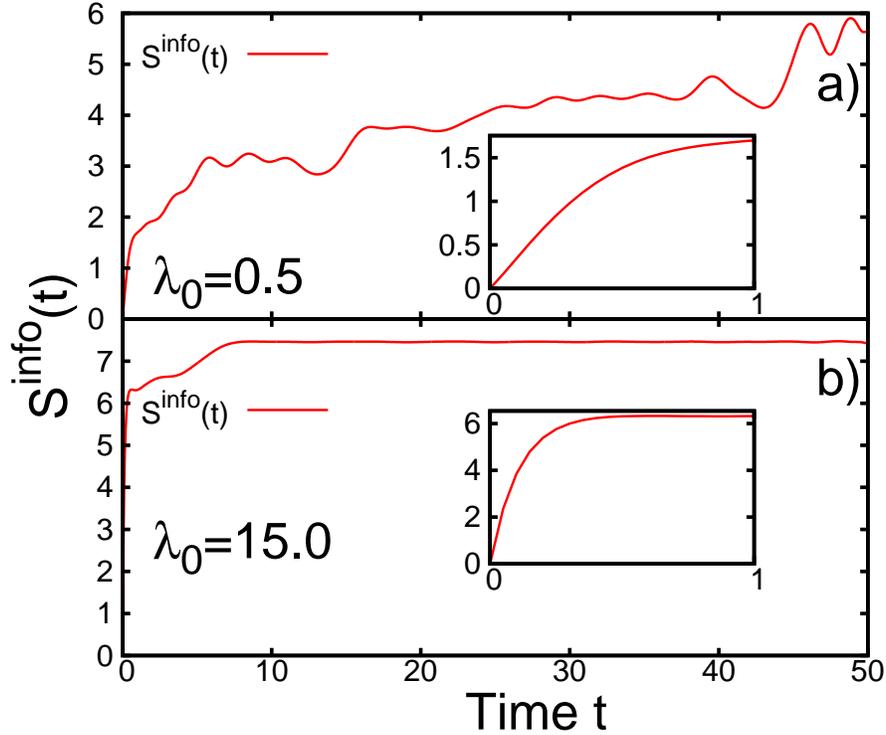}
  \end{center}

\caption{Long-time-evolution of many-body Shannon entropy $S^{info}(t)$ for weak
$\lambda_0=0.5$ and stronger $\lambda_0=15.0$ interactions. To verify that the
saturation of the many-body entropy measures is indeed non-transient, their
time-dependence for longer times is depicted. Panel a) shows $S^{info}(t)$ for
$\lambda_0=0.5$ and panel b) for $\lambda_0=15.0$. The behavior for smaller
times is presented in the insets. The saturation happens on a much faster
time-scale when the interaction strength is large with $\lambda_0=15.0$ as
compared to the situation when the interaction strength is small with
$\lambda_0=0.5$. See text for details, curves are smoothened for clarity of presentation, all quantities shown are dimensionless. }

\end{figure}

\begin{figure}
  \begin{center}
  \includegraphics[angle=0,width=10cm]{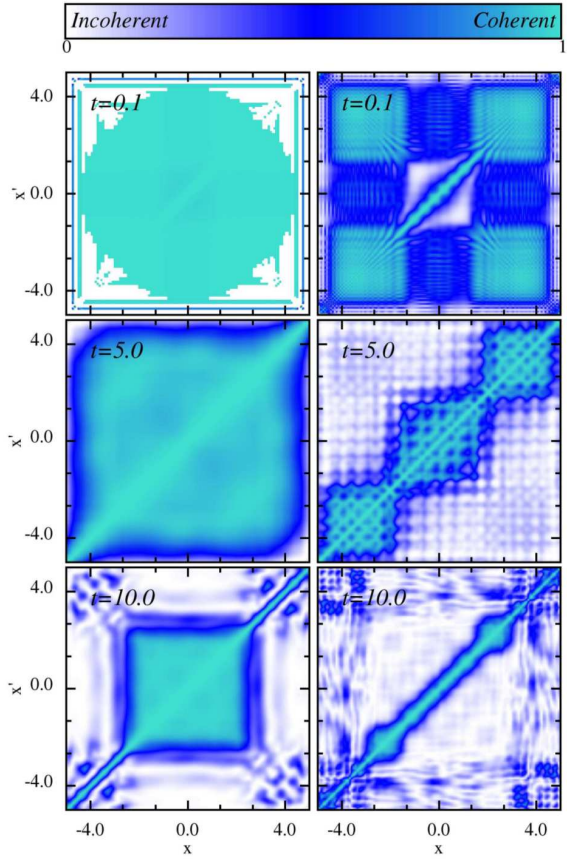}
  \end{center}
\caption{Coherence in the quench dynamics measured with the
correlation function $\vert g^{(1)}(x'_1, x_1 ; t)\vert^2$ for weak
$\lambda_0=0.5$ and strong $\lambda_0=10.0$ interaction strengths. The left
column depicts $\vert g^{(1)}\vert^2$ for $t=0.1,5.0,10.0$ with $\lambda_0=0.5$,
respectively. The right column shows $\vert g^{(1)}\vert^2$ for the same times
but for $\lambda_0=10.0$. The states with localized coefficient distributions and small entropies [compare Fig.~1 and Fig.~5a)] are also closer to being coherent ( $\vert g^{(1)}\vert^2 \approx 1$ ) than the states with large spread in coefficient distribution and large entropies [compare Fig.~2 and Figs.~5b) and 5c)].
In the case of spread-out coefficients and large many-body entropies $\vert g^{(1)}\vert^2 \approx 0$ holds almost everywhere but for the diagonal $\vert g^{(1)}(x,x;t) \vert^2$. Entropy production and loss of coherence hence go in-sync. See text for further discussion, all quantities shown are dimensionless.}
\end{figure}

\end{document}